# A Communication Protocol for Man-Machine Networks

Neda Hajiakhoond Bidoki
*Department of Computer Science*
*University of Central Florida*
Orlando, FL USA
nedahaji@cs.ucf.edu

Gita Sukthankar
*Department of Computer Science*
*University of Central Florida*
Orlando, FL USA
gitars@eecs.ucf.edu

*Abstract*—One of the most challenging coordination problems in artificial intelligence is to achieve successful collaboration across large-scale heterogeneous systems that include Robots, Agents, and People (RAP). In the best case, these RAP systems are potentially capable of leveraging the strengths of the individual entities to achieve complex distributed tasks. However, without intelligent communication protocols, man-machine partnerships are likely to fail as the humans become overloaded with irrelevant information. This paper introduces a communication protocol for man machine systems and demonstrates that its message routing performance approaches the central optimized solution in a simulated smart environment scenario.

## I. INTRODUCTION

The potential of man-machine teams has tantalized researchers for over a decade. Scerri et al. wrote a seminal paper introducing the acronym, RAP, to describe systems consisting of Robots Agents and People [1]. RAP systems leverage the strengths of the heterogeneous components, drawing from the common sense knowledge of the human, the robots' ability to perform repetitive physical tasks, and the ability of software agents to solve specialized artificial intelligence problems. They augment large-scale participatory sensor networks composed of humans carrying mobile devices with additional autonomous robot and software agents. Scerri et al. envisioned an architecture in which software agent proxies running on mobile devices could be used to coordinate the man-machine teams.

RAP systems are valuable for a variety of problems, including command and control, sensor networks, urban rescue, and personal assistance.

In the future, RAP systems are likely to become an integral part of smart cities, serving the function of proactively helping humans in urban areas. There have been limited demonstrations of HRI (human-robot interaction systems) as museum tour guides [2] and as building receptionists [3]. To extend these systems to include multiple robots and humans requires solving coordinated task allocation and scheduling which are NP-hard problems [4].

Glas et al. [5] introduced a general framework for networked robots that supports different social robot services including the observation of human behavior using environmental sensor networks, structured knowledge sharing, centralized resource and service allocation, global path planning for coordination between robots, and support for selected recognition and decision tasks by a human operator. In this paper, we propose new communication protocols to support this type of man machine system that includes networked robots cooperating with humans. We demonstrate that our new communication protocols are valuable for reducing communication costs in a simulated Netlogo scenario inspired by the Glas et al. shopping assistance system.

## II. PROBLEM STATEMENT

In a man-machine team, humans, robots, and agents must cooperate to achieve the joint goal. In our smart environment shopping assistance problem, customers are aided by a combination of fellow shoppers, mobile robots, and software agents who help locate a desired set of items. We define the problem as consisting of an environment defined by a map, a set of robots, $R$, a set of human-service assistants, $H$, a set of customers $C$, and a set of additional system constraints. The set of robots on the team is defined as $R := \{r_1, r_2, ..., r_N\}$ where $N$ is the number of robots on the team. The set of human service assistants on the team is defined as $H := \{h_1, h_2, ..., h_M\}$ where $M$ is the number of human service assistants on the team. To make it simpler, we consider one customer or request sender as c.

Our communication protocol must connect the customer with the best set of RAP assistants such that both human and non-human agents assist the customer to accomplish his requirements while minimizing cost. Finding a balance between reward and recruitment effort remains a challenge. We assume that humans' willingness to collaborate changes over time, and that the customer uses a monetary offer or tips to motivate other humans to provide assistance. According to the incentive theory of motivation, if people receive a positive profit from performing a task, there is a higher chance that they will successfully complete it. An analysis of workplace incentive programs suggests that correctly employed incentives are able to enhance participants' performance [6], [7] [8]. Additionally, prompt awards enhance participants' motivation even more; an instant award, combined with repetitive actions, can create new behavioral habits.

## III. METHODOLOGY

Our proposed protocol (History-based Financial Incentive) leverages the history of incentive acceptance to determine the best message routing. We compare our protocol against a centralized optimization algorithm to calculate the best possible agent allocation as well as Directed Diffusion protocol which is designed for robustness, scaling and energy efficiency in wireless sensor networks .

Centralized optimization algorithms are undesirable for RAP systems since they rely on centralized information as well as a single computational node. These characteristics reduce their ability to deal with large scale problems and datasets, due to the high computational complexity. Moreover, it is inefficient to collect and store data in a centralized manner especially when the communication is multicast. In such scenarios, collecting all the necessary information through a central node is both time-consuming and incurs a high communication cost due to the large amount of packet exchange. Having a single point of failure also jeopardizes the inherently resilient nature of RAP systems.

As an alternative to the centralized optimization algorithm, we implemented two protocols: The first is a Directed Diffusion algorithm and the second is our proposed history-based algorithm that tracks successful assistants.

- Directed Diffusion protocol

The customer requests as interests for named resources. Agents satisfying the interest can be found by flooding the message. Confirmation is exchanged by intermediate agents and resources are shipped when confirmed.

- History-based Financial Incentive protocol (**HFI**)

In our proposed protocol, the history of previously successful assists is recorded. Customers can make requests to the set of agents who are stored in his records. We believe that this protocol can reduce communication cost in many applications, especially when customers make repeated requests for similar types of assistance.

We have implemented these two algorithms along with the incentive strategy used for motivating human agents.

*A. Human Behavior Modeling*

To account for differences between the agents and humans, we created a separate human behavior model. Skill is often a major determinant of human success, yet it can be sensitive to situational factors. In human-robot interaction tasks requiring physical control, human skills have been found to change over both the short and long term. Over the long term body movements slowed down and/or became less accurate; simple control skills may exhibit different kinematics and dynamics and are affected by microgravity [9]. Also, time-of-day affects the individual's performance; for instance, circadian rhythms such as morningness or eveningness can impact productivity. Creating a physically realistic human behavior model is complex and beyond the scope of our work.

Instead in our scenario, we assume that skill is a negligible factor but that the human's current ability to perform the task can be modeled by a normal distribution, centered on the human's preferred time of day; this preferred time is when they are most available to render assistance. Peak time differs for each individual in our simulated scenario. Although robot performance can also fluctuate over time due to causes such as mechanical malfunctions and improper maintenance, we do not expect these situations to occur frequently in the short term.

Thus, we assume that the non-human agents use a greedy task acceptance model; whenever a robot is not busy with other tasks, it always renders assistance.

*B. Mathematical Model*

We present a mathematical model for the problem that is used to calculate the optimal allocation that serves as our comparison benchmark. $H$, $R$ and $C$ represent the set of human agents, robot agents and customers (and their locations) respectively. $h_i$, $r_i$ refers to $i^{th}$ human and $i^{th}$ robot agents. For simplicity, we consider one customer represented as $c$. Variables and parameters are as follows:

**Variables**
- $l_{h_i}$ : Binary variable assuming the value 1 if human agent $i$ has been selected to assist the customer; 0 otherwise.
- $o_{h_i}$ : Binary variable assuming the value 1 if human agent $i$ has been selected to assist the customer; 0 otherwise.
- $l_{r_i}$ : Binary variable assuming the value 1 if robot agent $i$ has been selected to assist the customer; 0 otherwise.
- $c_{h_i}$ : Number of human agents have been requested by current customer.
- $c_{r_i}$ : Number of robot agents have been requested by current customer.
- $c_t$: Current customer monetary offer.

**Parameters**
- $N_h$ and $H_h$: are the size of human, robot, agents sets respectively.

We assume each robot agent, human agent as well as customer is equipped with an IoT device with an omni-directional halfduplex antenna [10], [7], [11].

*C. Integer Linear Programming Formulation*

The objective function minimizes the communication cost of agents and customers through the process of customer agent resource assembling. For the robot costs we consider the shortest path the robot can take to reach the customer. For human agents we also include the reward costs required to motivate response. $E_{h_i}$ and $E_{r_i}$ represent the total cost for human agent $i$ and robot agent $i$ necessary to reach current customer and assist with his demand. Therefore the objective function can be written as follows:

$$\text{minimize } E = \sum_{h=1} l_{h_i} E_{h_i} + \sum_{r=1} l_{r_i} E_{r_i} \quad (1)$$

subject to the following constraints:

$$\sum_{h=1} l_{h_i} = c_i \text{ and } \sum_{r=1} l_{r_i} = c_{r_i} \quad (2)$$

$$l_{h_i} * o_{h_i} <= c_t \quad (3)$$

The first constraint ensures that the total number of human and robot agents that are selected is equal to the number of human and robot agents that the customer has requested. The second constraint guarantees that the minimum offer value of the selected human is less than what the customer has offered, thus ensuring that the human agent is motivated to assist the customer.

*D. Implementation*

We implemented our shopping assistance scenario and routing protocols in Netlogo which is a multi-agent programmable modeling environment [12]-[13]. Fig. 1 shows the simulator interface, and Fig. 2 shows a flow chart of its operation.

All variables are set when the simulation is initialized. Then a random customer creates a message with the required resource and reward info. This message is either broadcast to all surrounding agents (the flooding protocol) or unicast to the agents who have assisted the customer with previous demands. After a response, the message will be updated with the remaining required resources. If any required resources exist, the message is broadcast. This process will continue until no resources are needed or all the agents have been contacted. In the latter case, the customer can increase the reward to attract more help, and the process will repeat.

**Algorithm 1** History-based routing algorithm

Result: Message routing
Contact agents in history
  Update required resources
  **If** more resources are needed **then**
    **While** more resource needed or all agents have not received message **do**
      Upon receiving a message from an agent forward it to other nodes
      **if** an agent exhibits interest **then**
        Update the needed resources
        Forward message
      **else**
        Forward message
      **end**
    **end**
**end**

### IV. RESULTS

The results for the centralized optimization model were obtained by solving the ILP model using AIMMS run on a Windows-based 64-bit core-i7 computer with 24GB of RAM. In all the scenarios we considered, LP model executions were fast, never lasting more than few seconds. We implemented the heuristics in a home-grown software framework written in NetLogo. Their executions were similar, lasting only a few minutes according to the number of agents in the simulated environment. After initializing the map, we determine the number of agents (including the number of human agents, robot agents and customers) as well as our budget and the maximum number of human and robot agents that a customer can request as inputs. Our simulation then randomly places humans, robots, and customers on the map of a building, assuming constant sensor radio coverage. For each scenario created in the NetLogo simulation, we run the two communication protocols (DD and HFI). The map and model info were loaded directly into our AIMMS program in order to execute the optimization procedure. In this way we are able to calculate the results of all three models on the same scenario.

Initialization parameters (including number of agents) were varied; for each set of parameters, 20 different scenarios were generated. The average cost of all 20 different scenarios is used as the communication cost. Fig. 3 shows the results; as expected the History-based Financial Incentive algorithm (marked as HFI) had a better performance, closely matching the optimum selection of agents. This occurs due to several facts. First agents who previously participated on a team are likely to be around, having recently finished their previous task.

The DD protocol does not perform well in comparison to the HFI algorithm as it constantly broadcasts to the surrounding agents. Obstacles such as walls do not block signal reception of agents but do block movement. Therefore, an agent may receive a request quickly while needing take a long path in order to reach the customer, due to the existence of solid obstacles.

A matched paired t-test on the mean values of the same scenarios under different algorithms yields no significant difference between our proposed FIH protocol and the optimal ILP solution. Comparing DD protocol communication cost with HIF communication cost yielded t equal to -7.937337, indicating that the result is significant at $p \leq 0.01$.

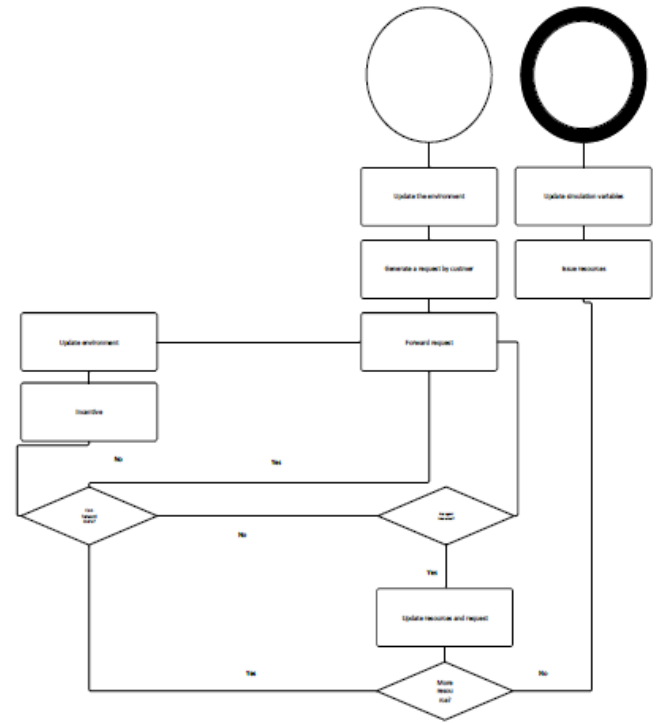

Fig. 2. Simulation logic flow chart

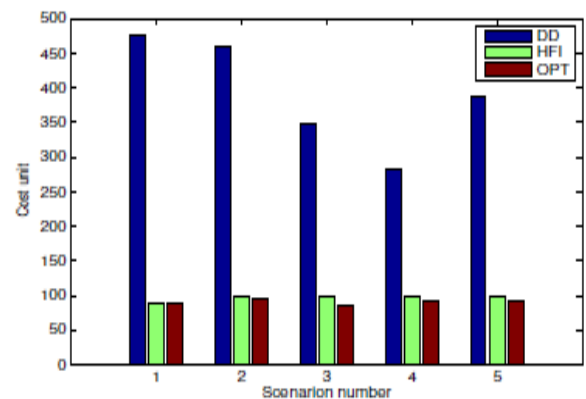

Fig. 3. Communication cost for each routing protocol

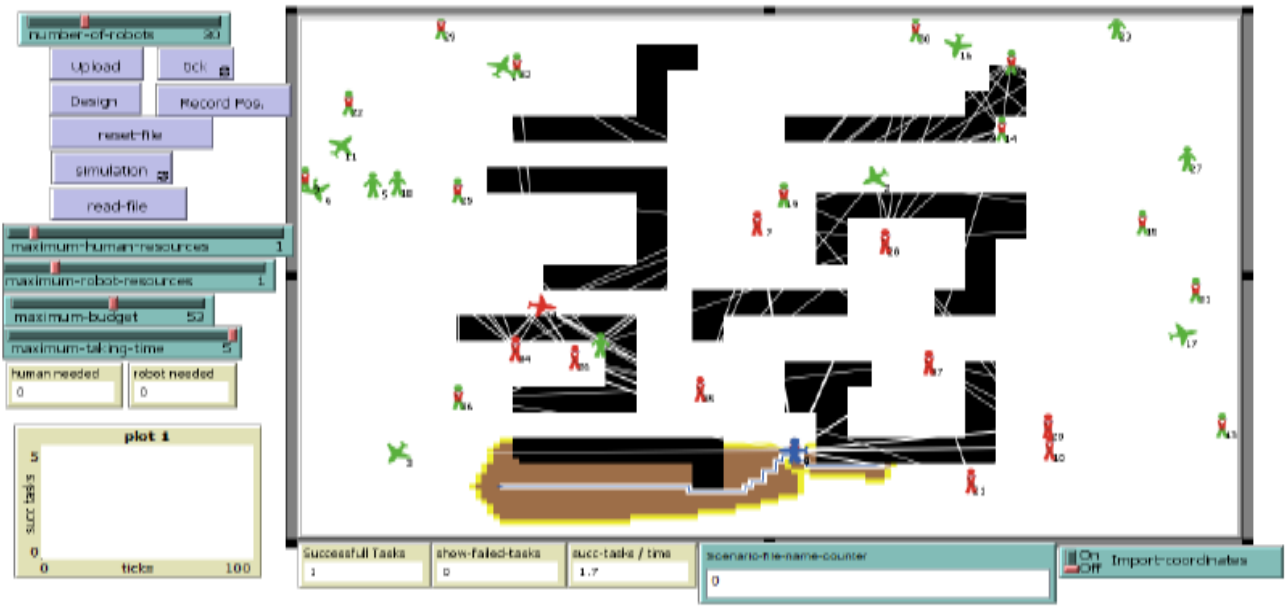

Fig. 1. Netlogo simulation of shopping assistance scenario

## V. CONCLUSION

This paper introduces a history-based financial incentive communication algorithm for man-machine systems. NetLogo was used to simulate a shopping assistance scenario in which a smart store environment summons help for the shopper in the form of robots and other humans to help locate items. Although non-human agents have no reason not to respond if available, humans are likely to be performing other shopping tasks and need to be incentivized to render assistance. In our simulation, they are modeled as having time availability preferences and as being less willing to respond to lower incentives outside their peak availability period. We demonstrate that the agent allocation solution reached our proposed algorithm results in an insignificant cost increase over a centralized solution calculated with an ILP solver. Lower communication costs are particularly important in man-machine systems to avoid bombarding the human with unwanted messages. In future work, we plan to implement our communication protocol in ROS (the Robot Operating System) for use in coordinating quadcopters with humans for autonomous photography tasks.